\documentclass[sigplan,10pt]{acmart}
\setcopyright{none}

\AtBeginDocument{%
  }

\usepackage{amssymb}
\usepackage{amsfonts}
\usepackage{amsmath}
\usepackage{tikz}
\usepackage{amsmath}
\usepackage{algorithm}
\usepackage[noend]{algpseudocode}  

\usepackage{filecontents}
\usepackage{circuitikz}
\usepackage{makecell}  
\usepackage{circledsteps}
\usepackage{subcaption}
\usepackage{booktabs}
\usepackage{pifont}
\usepackage{xspace}

\usepackage{array}
\newcolumntype{C}[1]{>{\centering\arraybackslash}p{#1}}
\newcommand{\name}{EasyRider\xspace}

\begin{document}
\date{}

\title{EasyRider: Mitigating Power Transients in Datacenter-Scale Training Workloads}

\author{
{\rm Dillon Jensen}\\
\and
{\rm Grant Wilkins}\\
\and
{\rm Obi Nnorom Jr.}\\
\and
{\rm Hugo Budd}\\
\and
{\rm Ram Rajagopal}\\
\and
{\rm Juan Rivas-Davila}\\
\and
{\rm Phil Levis}\\
\textbf{Stanford University}
}

\begin{abstract}
Large-scale AI model training workloads use thousands of GPUs
	operating in tightly synchronized loops. During synchronous communication,
	start-up, shut-down, and checkpointing, 
	GPU power consumption can swing from peak to idle 
	within milliseconds. Such steep power ramp rates induce reactive power transients, leading to voltage and frequency shifts that can damage transformers, generators, and protection equipment on the broader power grid.

To solve this problem, we introduce \name, a power delivery architecture to
	mitigate power fluctuations at the rack level.
	\name~ uses passive and active hardware components 
	to attenuate high-frequency power swings and
	rack-scale energy storage for the large amounts of energy
	needed to smooth high-power racks. A software system
	continually monitors the energy storage system to maximize its
	lifetime in the presence of frequent charge/discharge cycles.
	\name filters rack power variations to stay within grid-specified
	limits without requiring software
	modifications to AI training frameworks or wasting 
	energy. We evaluate \name on a 10 kW/400 Vdc-rated
	rack-scale prototype,
	demonstrating its effectiveness across heterogeneous power levels and 
	workload power profiles.
\end{abstract}

\maketitle
\pagestyle{plain}
\section{Introduction}

\begin{figure}
\centering
\includegraphics[
    width=0.9\columnwidth,
    trim=0.25cm 0.25cm 0.25cm 0.25cm,
    clip]
    {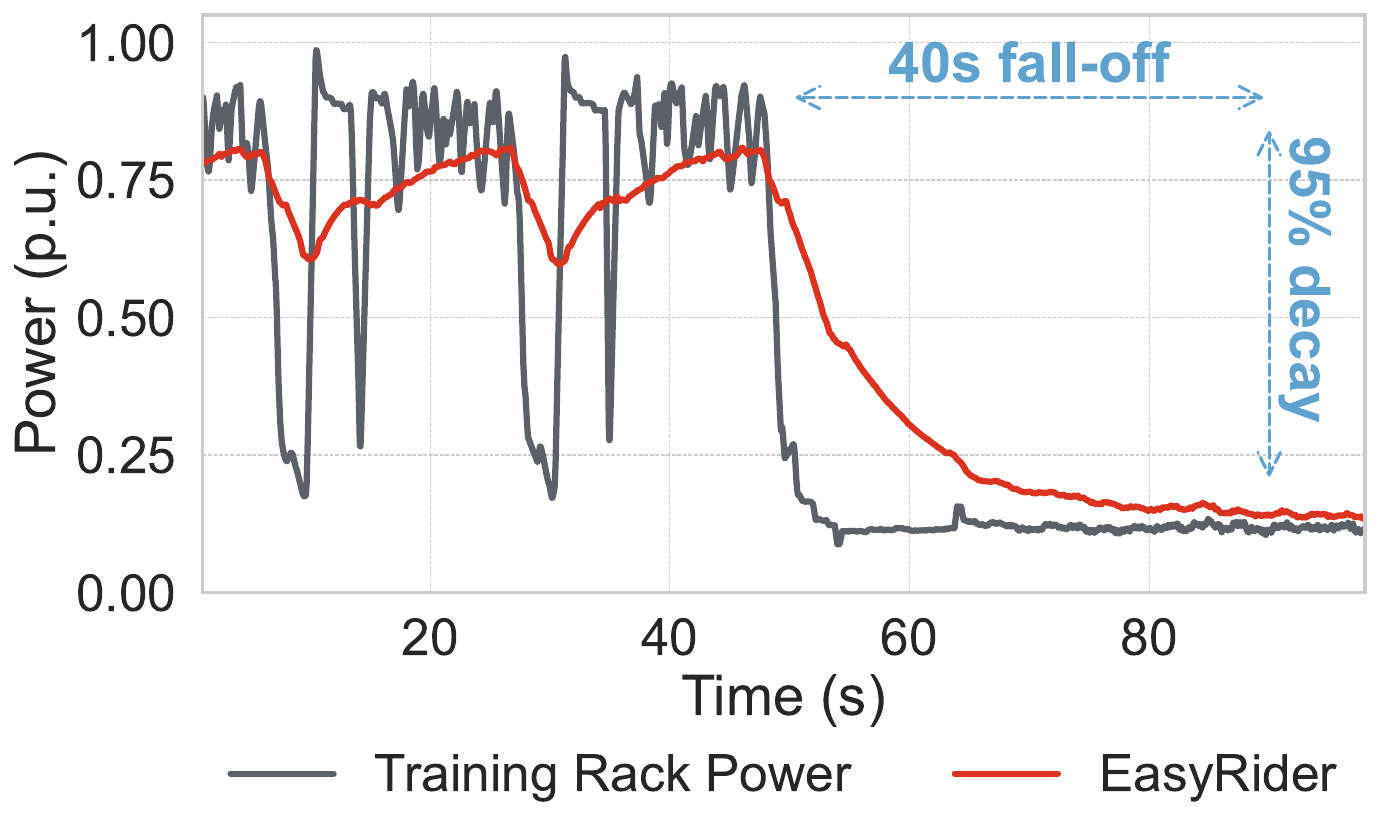}
\caption{The \name prototype is able to smooth the rack power draw to within grid ramp rate limits. While rack power drops rapidly by 80\%, the grid observes a gradual power draw change over tens of seconds.}
\label{fig:demonstration}
\end{figure}

Pre-training foundation models such as OpenAI's GPT series, Anthropic's Claude,
or Google's Gemini require tens or hundreds of thousands of GPUs/TPUs.  During
training, these accelerators execute synchronous compute-communicate iterations:
in near lockstep, they all compute, pause to communicate, then compute again
\cite{li2025ailoaddynamicsapower, li2024unseenaidisruptionspower}. During
communication events, the power draw of all of the accelerators swings between
maximum and idle power within milliseconds; for an NVIDIA H100, for example, this
is a near-instantaneous 80\% power
reduction, as shown in Figure~\ref{fig:demonstration}~\cite{semianalysis2025,microsoft_power_stabilization}. A modern training job that
uses 50,000 GPUs (e.g. Meta's Llama-3~\cite{llama3}) might draw 35MW when computing
and drop to 7MW during communication.

The fundamental challenge with such swings is that the stability of the power
grid requires a constant balance between supply and demand.
To match supply to demand, the grid relies on generators that adjust 
output in response to load changes. Because generators are 
physical systems, they are bound by mechanical limits on 
how quickly they can adjust their output.
When power demand changes faster than generators can respond, 
the grid's frequency and voltage move outside their narrow safe
ranges, risking damage to generators and other connected
equipment~\cite{kirby2003frequency}.


The grid protects itself from damage by
automatically disconnecting unstable loads~\cite{disconnect}. However, when those
loads are multi-gigawatt datacenters, the sudden loss of demand is itself a
destabilizing event. The possibility of grid damage and blackouts from training loads has become a
major impediment to building new datacenters. In some recent cases, new datacenter
projects have been denied because of the instability that training can bring to
the grid~\cite{datacenter-denied}.

This paper proposes~\name, a novel rack-level power supply architecture
which allows an AI-training rack to easily "ride through"
any power transient.
From the perspective of the grid, large power swings on an~\name-equipped rack appear as a gradually changing power draw. 
Figure~\ref{fig:demonstration} demonstrates this 
automatic power-smoothing behavior, with millisecond-scale ramping events in a commercial rack power trace being smoothed over thirty seconds or more.
Designed for compatibility with the future $400$ or $800 V_{DC}$
datacenter regime, \name~can be easily configured to satisfy
any grid-imposed ramp-rate restrictions without requiring changes to existing
software, models, or GPU firmware, and without requiring additional hardware systems at the datacenter or grid scale.


This paper makes three main research contributions:

\vspace{1ex}
\noindent{\bf{Hardware system design and testing:}} \name~simultaneously 
powers a rack and removes power transients through a 
combination of passive and regulated (active) subsystems. 
Passive components (capacitors and inductors) filter 
high-frequency events, and high-power batteries are used in 
closed-loop control to filter longer, low-frequency power 
fluctuations. Smoothing transients in hardware has numerous 
other advantages: the system does not depend on software for 
measurement and control, and can therefore respond reliably 
and with minimal latency. It can be engineered to tolerate an 
arbitrary load up to a given maximum power magnitude.

We show how this effect scales at the cluster level, ensuring that the entire datacenter will stay compliant with grid-imposed restrictions on load behavior.

\vspace{1ex}
\noindent{\bf{Auxiliary software design and testing:}} \name~uses an optimization-based control loop to dynamically regulate the state of charge (SoC) for the battery system. Tracking a target SoC ensures EasyRider will have sufficient stored energy to smooth future transients despite charging and discharging inefficiencies. This additionally extends battery life by avoiding deep discharges or over-charging.

\vspace{1ex}
\noindent{\bf{Scaling and lifetime analysis:}} We analyze the practicality of \name~as it might be deployed in real-world systems, elucidating the smoothing effect at the cluster- or datacenter-wide scale. We explore the expected battery lifetime using state-of-the-art simulation, estimating system costs and expected energy savings compared to other proposed solutions.
\section{Background and Motivation}
\label{sec:background}

This section provides background on the topics that motivate \name's design. First, it describes how the grid delivers power to datacenters and why AI-training loads change so much quicker than traditional cloud-computing power consumption. We also explain typical power architecture in modern datacenters as well as some current and proposed approaches to protect the grid from training transients.

\subsection{Grid Operating Parameters}\label{sec:background:grid}
At any given moment in time, supply and demand for power in the grid are equal across the aggregate of all connected generators and loads. Some generators, such as solar panels and wind turbines, produce power according to the weather; others, such as gas, coal, nuclear, and hydro, can be controlled dynamically. The rate at which a generator can increase or decrease its power output is called the {\it ramp rate}~\cite{NERC2021_oscillations}. Similar to how a car cannot instantly go from idle to its top speed, a turbine generator cannot instantly change its electrical output. Depending on the 
generator, ramping times (i.e. from idle to peak power output or vice-versa)
can range from a few seconds to
several hours, and bringing new generators online can take minutes to days
\cite{GONZALEZSALAZAR20181497, ramp-rates-260859}. For high-ramp rate gas turbines, the generator's output might be able to change by 10-20~MW/min, requiring several minutes to ramp from idle to its maximum output.~\cite{NERC2021_oscillations,nerc2025largeloads,maximum-ramp-rate}. These ramp rates have historically been sufficient because the aggregate load in the grid changes slowly.

At the same time, loads make assumptions about generators: in the United States, for example, the grid provides residential power at $120V_{RMS}$ and 60Hz, but this can vary, and must remain within 114\--126$V_{RMS}$ and 59.9--60.1Hz~\cite{AESO2025datacentreconnection,nerc2025largeloads}. Devices attached to the grid assume these conditions and can be damaged if power moves outside these ranges~\cite{NERC2021_oscillations}.

If aggregate load changes faster than generators can adapt, the grid voltage and/or frequency will rise or sag in response~\cite{NERC2021_oscillations}.
The danger and damage of large, fast swings is real:
A July 2024
transmission fault caused 1.5 GW of datacenters in Northern Virginia to
simultaneously disconnect, requiring emergency grid management to prevent
widespread outages~\cite{nerc2024}.
The Electric Reliability Council of Texas (ERCOT) has
identified that simultaneous disconnection of 2.0--2.6 GW of datacenter load
could destabilize the entire Texas grid and trigger cascading
blackouts~\cite{ercot2025}.

Grids and generators can also be susceptible to damage caused by loads that excite resonant frequencies in different hardware systems: NERC's analysis of a 2019
disturbance showed that a load oscillating at 0.25~Hz propagated across
interconnections and damaged generators hundreds of miles
away~\cite{nerc2019oscillation,GE_torsional_dynamics}.

\subsection{AI/ML Training Power Dynamics} 

Historically, datacenters have operated within grid operating parameters.
Cloud-computing datacenters run tens or hundreds of millions of different jobs,
each of which is a tiny load; control planes such as Kubernetes~\cite{kubernetes}
or Borg~\cite{borg} stagger job starts over seconds, and the aggregate power use changes slowly over minutes or hours.

Large-scale distributed training behaves differently.
During synchronous data-parallel training, clusters of
tens or hundreds of thousands of GPUs execute in lockstep: 
they compute gradients locally, then
synchronize parameters through collective communication primitives like
exchanging tensors or gradients.

This lockstep execution creates large, sudden steps in power that are observed at the cluster level.
Large clusters using modern accelerators may shed 80-95\% of their load in fractions of a second~\cite{li2024unseenaidisruptionspower,nvidia-caps,microsoft_power_stabilization,
semianalysis2025}. These are not rare events---they occur every few seconds at the end of a training
iteration, and (often unpredictably) every few minutes during checkpointing, restart, or
a fault or collective stall.


\subsection{Datacenter Power Hierarchy}


\begin{figure}
    \centering
    \includegraphics[width=0.8\columnwidth]{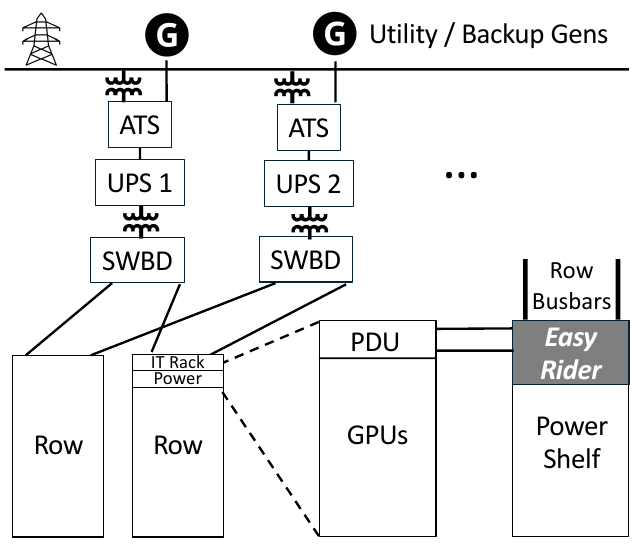}
    \caption{Modern data center power hierarchy and where \name~fits in. 
    This particular design shows disaggregated power from the rack with a connection to busbar distribution per-row. 
    Design variations may include in-rack UPSes or other power conversion components.}
    \label{fig:lineup-fig}
\end{figure}

Figure~\ref{fig:lineup-fig} shows how modern hyperscale data centers use 
multi-stage
power-delivery systems that transform power from high-voltage 13.8~kV utility
interfaces\footnote{Utilities consider 13.8kV a ``medium'' voltage, with
$>$100kV, e.g., for long-range transmission lines being ``high' voltage.} 
down to the 0.8-1.2~V low voltages at the processor level.
Utility power enters the datacenter and an on-site substation steps it down
to medium- and low-voltage switchgear.
Power is then routed through uninterruptible power supplies (UPSes), floor- or
row-level power distribution units (PDUs), and branch circuits to individual
racks~\cite{datacenter_as_computer,hamilton,zhang2021flex,hsu2018smoothoperator,piga2024dvfs}.

Datacenters have historically delivered ~48 V DC power to racks, with each CPU-centric rack using an aggregate peak power on the order of 30 kW~\cite{datacenter_as_computer}. Advances in power electronics and a desire for low networking latency in synchronous training have driven the development of increasingly power-dense racks, with modern AI accelerators drawing $>$100~kW and future-facing standards paving the way for 1~MW power sidecars, with racks or rows operating on ±400 $V_{DC}$ or 800 $V_{DC}$ busbars for improved electrical efficiency~\cite{nvidia-blackwell-tdp,kyber,kyber2,mtdiablo, ocp2026lvdc_2,pratt2007evaluation,ocp2026lvdc_1,nvidia2026_800vdc}. Although power regulation circuits smooth transients from individual <1 ms processor events, and conventional UPS systems will have energy stored to temporarily power racks in the event of a grid-side fault, these systems are not designed to smooth seconds-long rack-side power transients.

\subsection{Existing Approaches}
\label{sec:background:approaches}
\begin{table*}
	\centering
	\caption{Transient mitigation approaches. The key distinction is where mitigation is inserted and whether the high frequency transients are electrically or software-mediated.}
	\label{tab:approach-comparison}
	\setlength{\tabcolsep}{4pt}
	\resizebox{\linewidth}{!}{
	\begin{tabular}{llllll}
	  \toprule
	  \textbf{Approach} & \textbf{Placement} & \textbf{High frequency} & \textbf{Low frequency} & \textbf{SW/FW dependence} & \textbf{Main limitation} \\
	  \midrule
	  GPU burn~\cite{semianalysis2025} & GPU & None & Work injection & Training stack & Energy waste; no hardware protection \\
	  GB300 support~\cite{nvidia-caps} & Power shelf & Capacitors & Power cap / burn & Platform firmware & Energy waste; platform-specific \\
	  Software-controlled batteries~\cite{microsoft_power_stabilization} & Rack & Battery, SW-triggered & Battery + cap + burn & Telemetry + software & Fast path limited by telemetry \\
	  Site-level BESS~\cite{xai-power-stabilization} & Substation & None & Site battery & Site controller & Expensive, does not protect internal DC distribution \\
	  \midrule
	  \textbf{\name (ours)} & Rack or row PDU & Passive LC & Local battery & None for transient mitigation & Rack-local only \\
	  \bottomrule
	\end{tabular}}
  \end{table*}
A training cluster is ``grid-safe'' only if its load swings are attenuated 
before they reach the upstream electrical system. The relevant dynamics 
span both slower events such as job transitions and checkpoints and
faster content from processor-level dynamics. Existing mitigations address parts of 
this problem, but at different points in the hierarchy and with different 
dependencies on software, firmware, or 
site infrastructure~\cite{nerc2019oscillation,microsoft_power_stabilization,nvidia-caps,xai-power-stabilization}. 
Table~\ref{tab:approach-comparison} summarizes this design space.

\noindent\textbf{Software ``burn'' or scheduler-based smoothing.}
One approach is to inject secondary work, such as GEMM kernels, to artificially boost 
GPU activity when power falls below a target~\cite{microsoft_power_stabilization}. 
This can smooth some utilization drops, but only by spending extra energy and 
coupling protection to the training stack. If detection or control fails, 
the transient is exposed upstream.

Similarly, bubble-filling and high-utilization schedulers can reduce some iteration-level
swings by keeping GPUs more uniformly 
utilized~\cite{pipefill,deepseekai2025deepseekv3technicalreport,zerobubble,pipemorph}. 
These methods are complementary, but they do not provide an electrical guarantee 
at the rack boundary and remain sensitive to workload structure, checkpointing, and unpredictable fault/recovery events.

\noindent\textbf{Platform-specific electrical support.}
NVIDIA's GB300 NVL72 adds power-shelf capacitors together with startup power 
capping and ramp-down support~\cite{nvidia-caps}. This provides electrical 
mitigation for short ($\leq$ 60 ms) transients, but it is tied to a specific platform and 
does not address larger energy imbalances over longer events.

\noindent\textbf{Software-coordinated rack-level energy storage.}
Another approach is to combine rack-level batteries that dispatch
on software-triggered events along with GPU power capping like the NVIDIA
GB300~\cite{nvidia-caps,microsoft_power_stabilization}. However, this approach is not fault-tolerant, as battery discharge needs to be triggered by software controllers---it can take up to a second for rack telemetry to detect that a power ramping event has occurred, in which time the opportunity to smooth the transient has largely been missed~\cite{dimitrov2025gb300powersmoothing,saifhashemi2026gpupowersmoothing}. Any fault in the software stack could prevent the system from responding to a transient altogether.

\noindent\textbf{Site-level BESS.}
Site batteries buffer the aggregate load seen at the grid interconnection 
point~\cite{xai-power-stabilization}. This helps with slower site-wide variation, but comes with three key downsides: first, because the BESS sits above the internal row and rack distribution hierarchy, it does not stop rack transients from propagating through the 
internal power distribution before they are absorbed at the site boundary.

More importantly, this solution promotes costly overbuilding of expensive battery capacity. A site-level BESS requires provisioning for the expected peak IT power use up front, whereas incremental deployment at the rack or row level scales naturally with evolving compute deployments.

Finally, centralizing power smoothing at the campus level creates a single point of failure for transient mitigation and therefore may pose a larger risk of introducing transients back to the grid.

\section{Problem Formulation}
\label{sec:problem-statement}

\begin{figure}
	\centering
	\begin{subfigure}{\columnwidth}
		\includegraphics[
        width=0.95\linewidth,
        trim=0.0 0.35cm 0.0cm 0.3cm,
        clip
    ]{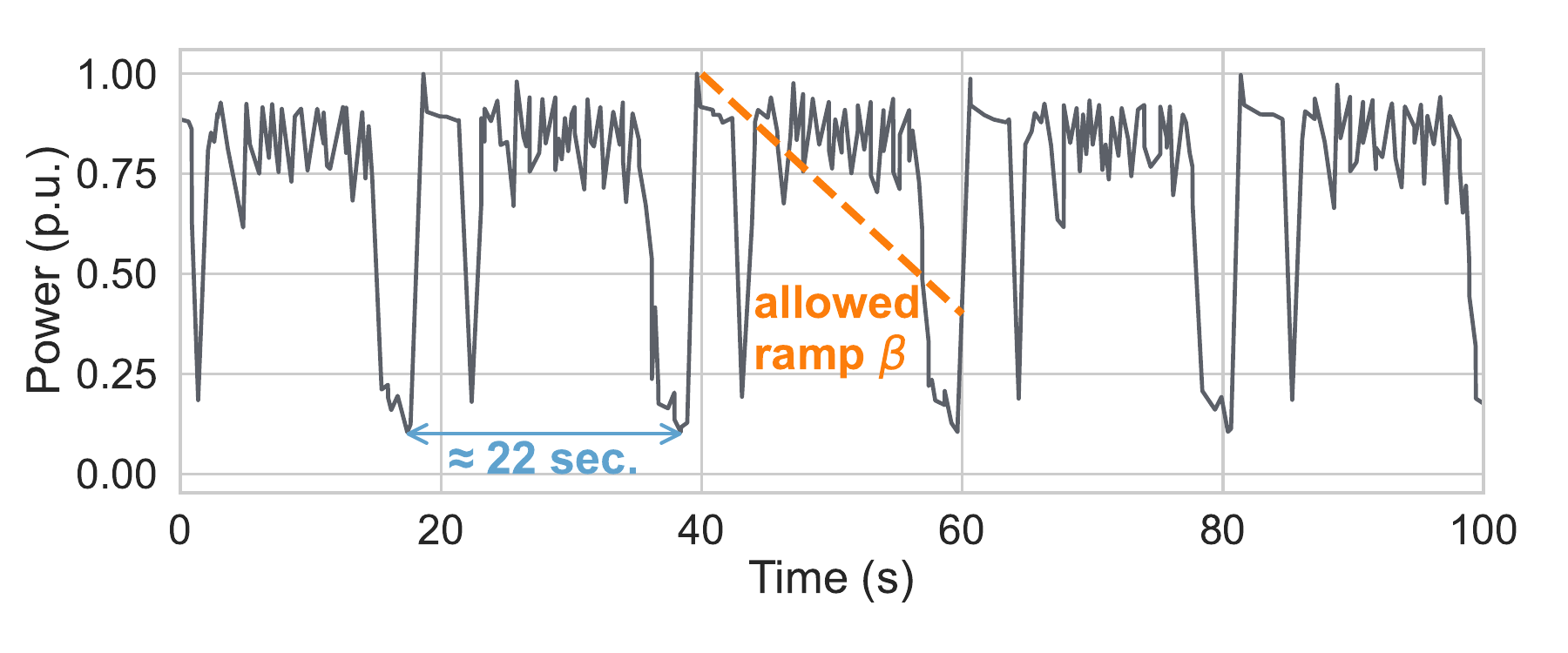}
		\caption{Time-domain.}
		\label{fig:microsoft_trace-time}
	\end{subfigure}
	\begin{subfigure}{\columnwidth}
		\includegraphics[
        width=0.95\linewidth,
        trim=0.0 0.35cm 0.0cm 0.3cm,
        clip
    ]{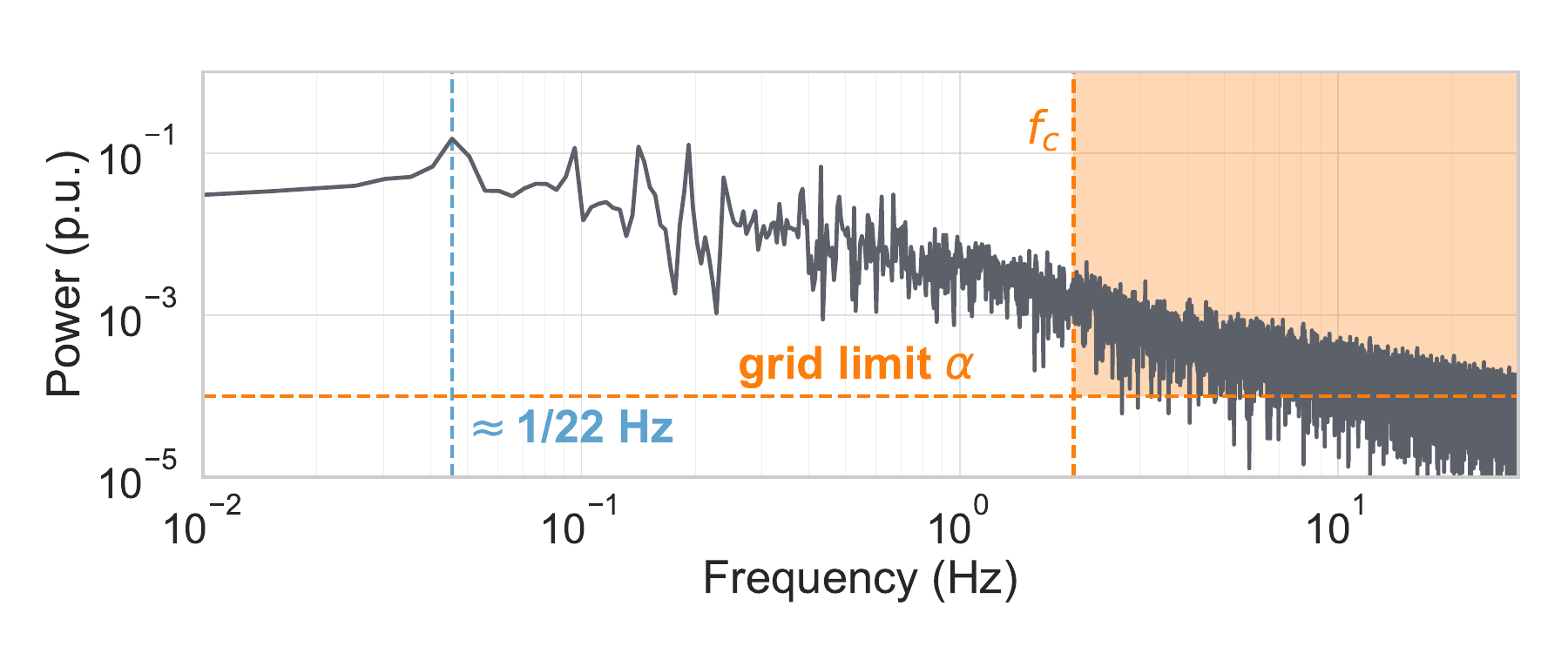}
		\caption{Frequency-domain.}
		\label{fig:microsoft_trace-freq}
	\end{subfigure}
	\caption{Time- and frequency-domain representation of a power trace based on Fig.~1 from~\cite{microsoft_power_stabilization}, which we use as a testbench for our \name prototype. The largest dips occur at approximately 22-second intervals, producing a prominent peak near $1/22$~Hz. $\beta$ is the allowed ramp rate by the grid operator and $f_c$ is the cutoff frequency for the grid limit $\alpha$.}
	\label{fig:microsoft_trace}
\end{figure}

The grid does not see individual GPUs, it only sees the aggregate power draw
of a datacenter and requires that this composite signal be ``well-behaved.''
As discussed in Section~\ref{sec:background:grid}, large-scale training
violates this expectation by creating large changes in power draw that occur
faster than generators and protection equipment can safely respond.

Grid operators are also concerned with the frequency content in a datacenter's power use: different generators and transmission line networks will be sensitive to resonant oscillations at frequencies ranging from less than 1 Hz to 100 Hz or so~\cite{nerc2019oscillation,NERC2021_oscillations,microsoft_power_stabilization}.

To make these constraints easy to reason about, we view the datacenter power
trace not just as a time series but as a sum of sinusoids at different
frequencies, obtained via a Discrete Fourier Transform (DFT). This frequency-domain view provides a useful way to characterize how rapidly a datacenter's power demand changes, and which frequencies need to be attenuated to protect upstream systems.

Let $P(t)$ denote the normalized time-domain power delivered to a rack or training cluster. Its DFT, $S(f)$, indicates how strongly the power trace varies at each frequency $f$: low frequencies capture slow changes in demand, while higher frequencies capture faster fluctuations.
Figure~\ref{fig:microsoft_trace} shows this, plotting the time-domain measured rack power and the corresponding magnitudes of the DFT. Events exhibiting strong time-domain periodicity with frequency \textit{f} will cause a high-magnitude spike at the same frequency in the DFT magnitude plot. For example, the trace shown in Figure~\ref{fig:microsoft_trace-time} repeats nearly identically every 22 seconds, leading to a peak in frequency content at $f=1/22 Hz$ in Figure~\ref{fig:microsoft_trace-freq}.



Grid operators generally impose two kinds of limits on $P(t)$ and $S(f)$:

\paragraph{Frequency content.}
The first limit constrains how much variation is allowed at high frequencies.
As a simple example, the grid operator might specify that all content above some cutoff frequency $f_c$ can have a magnitude of at most $\alpha$:
\begin{equation}
  \label{eq:frequency_content_spec}
  S(f) \le \alpha \qquad \text{for all } f \ge f_c.
\end{equation}
Above $f_c$, power fluctuations may need to be dampened in order to bring the system into compliance. Figure~\ref{fig:microsoft_trace-freq} shows
this constraint: the grey curve is $S(f)$ on a log–log scale, and any portion
above the horizontal line at $\alpha$ for $f \ge f_c$ violates the spec.

\paragraph{Maximum ramp rate.}
The second limit bounds how quickly the datacenter power can change in time:
\begin{equation}
  \label{eq:ramp_rate_spec}
  \left|\frac{dP}{dt}\right| \le \beta \qquad \text{for all } t,
\end{equation}
for some ramp-rate limit $\beta$ expressed as a fraction of rated power per
second.
Figure~\ref{fig:microsoft_trace-time} illustrates a
power trace that repeatedly exceeds this slope bound. 

Because both these specifications are written with respect to the total datacenter's peak load, an AI training job won't violate grid-imposed constraints if each rack in the cluster individually also complies with equations \ref{eq:frequency_content_spec} and \ref{eq:ramp_rate_spec}. This is expressed mathematically in Appendix~\ref{app:smoothing-at-scale}.
\section{\name Architecture}
\label{sec:architecture}
\begin{figure}
    \centering
    \includegraphics[width=0.9\columnwidth]{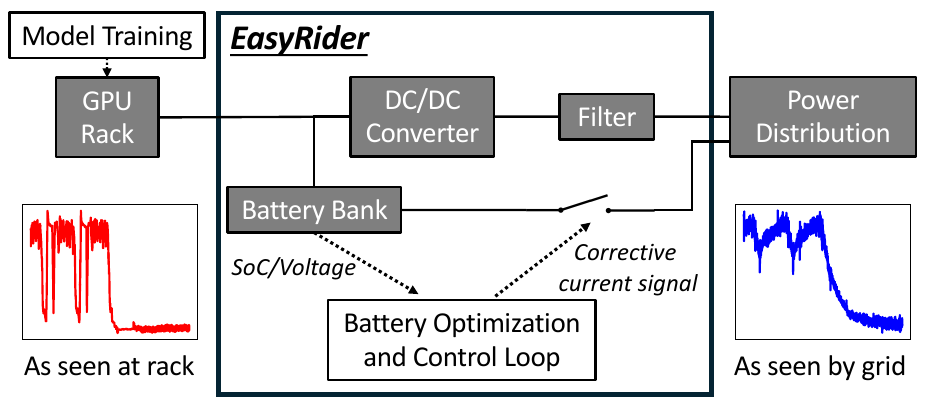}
    \caption{\name architecture. Software components are shown in white, hardware in gray. \name is agnostic to the training workload and can be integrated into existing datacenter power hierarchies with appropriate conversions and component sizing.}
    \label{fig:system-fig}
\end{figure}

\name~is a PDU system that sits between a GPU rack and the row bus, smoothing the rack's power
waveform before it reaches the rest of the datacenter power system. Its actions are
invisible to upstream devices: UPSes, PDUs, and substations all see a
grid-compliant, low-ramp rack load. Because it 
depends only on local sensing and actuation, the \name~architecture can be adapted for next-generation racks with power sidecars~\cite{kyber} or retrofitted into
existing datacenters with in-rack PDUs without changes to the cluster software
stack.

As illustrated in Figure~\ref{fig:system-fig}, \name~comprises three physical
elements plus controls and a software system, with a clean decomposition.
The hardware systems power the rack and smooth transient events. An auxiliary software system regulates battery SoC over longer timescales to maximize system lifetime.

Together, these stages present a smoothed rack load that satisfies
the frequency and ramp-rate specifications from
Section~\ref{sec:problem-statement} while leaving the underlying training job
unchanged. Section~\ref{sec:hardware-design} details the filter, converter, and
battery design; Section~\ref{sec:software-control} describes the control loop
that preserves battery SoC and lifetime.
\section{Hardware Design}
\label{sec:hardware-design}

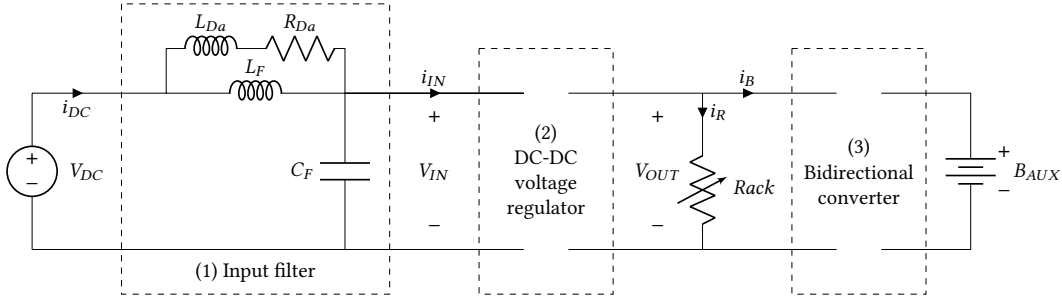
\begin{figure*}[!htbp]
    \begin{center}
    \resizebox{0.8\linewidth}{!}{
    \centering
    \begin{circuitikz}[american voltages]
    \def\fcScale{0.75}
    \def\border{1*\fcScale}
    \def\dampingLine{4.5*\fcScale}
    \def\topLine{3.5*\fcScale}
    \def\bottomLine{0*\fcScale}
    \def\VDCx{-2*\fcScale}
    \def\filterLeftx{1*\fcScale}
    \def\filterRightx{5*\fcScale}
    \def\dampingCenterx{3*\fcScale}
    \def\CFx{5*\fcScale}
        \draw (\VDCx,\topLine) to [V=$V_{DC}$] (\VDCx,\bottomLine) 
            (\VDCx, \topLine) to [short, i_=$i_{DC}$](\filterLeftx-\border,\topLine) -- (\filterLeftx,\topLine)
            to [L=$L_F$] (\filterRightx,\topLine) 
            (\VDCx,\bottomLine)-- (\filterRightx,\bottomLine) to [C={$C_F$}] (\filterRightx,\topLine) 
            (\filterLeftx,\topLine) -- (\filterLeftx,\dampingLine)
            to [L=$L_{Da}$] (\dampingCenterx,\dampingLine) 
            to [R=$R_{Da}$] (\filterRightx,\dampingLine) 
            -- (\filterRightx,\topLine);
        \draw[dashed]
            (\filterLeftx-\border,\dampingLine+\border) rectangle (\filterRightx+\border,\bottomLine-\border);
        \node at (\dampingCenterx,\bottomLine-\border/2) {(1) Input filter};
    \def\VregLeftx{9*\fcScale}
    \def\VregRightx{10*\fcScale}
    \def\GPUsx{13*\fcScale}
    \def\GPUsYTop{2}
        \draw (\CFx,\topLine) -- (\VregLeftx,\topLine)
        ({(\CFx+\VregLeftx)/2},\topLine) to [open, v=$V_{IN}$]({(\CFx+\VregLeftx)/2},\bottomLine) 
        (\CFx,\topLine) to [short, i=$i_{IN}$](\VregLeftx,\topLine) 
        (\CFx,\bottomLine) -- (\VregLeftx,\bottomLine)
        (\VregRightx,\topLine) -- (\GPUsx,\topLine)
        (\VregRightx,\bottomLine) -- (\GPUsx,\bottomLine)
        (\GPUsx,\topLine) to [short,i=$i_{R}$](\GPUsx,\GPUsYTop) to [vR=$Rack$, invert](\GPUsx,\bottomLine) 
        ({(\VregRightx+\border+\GPUsx)/2},\topLine) to [open, v=$V_{OUT}$]({(\VregRightx+\border+\GPUsx)/2},\bottomLine);
        \draw[dashed]
            (\VregLeftx-\border,\topLine+\border) rectangle
            (\VregRightx+\border,\bottomLine-\border);
        \node[align=center] at 
            ({(\VregLeftx+\VregRightx)/2}, {(\topLine+\bottomLine)/2}) 
            {(2)\\DC-DC\\voltage\\regulator};
    \def\BidirectionalLeftx{16*\fcScale}
    \def\BidirectionalRightx{17*\fcScale}
    \def\Batx{19*\fcScale}
        \draw (\GPUsx,\topLine) to [short, i=$i_{B}$](\BidirectionalLeftx-\border,\topLine) -- (\BidirectionalLeftx,\topLine)
            (\GPUsx,\bottomLine) -- (\BidirectionalLeftx,\bottomLine)
            (\BidirectionalRightx,\topLine) -- (\Batx,\topLine)
            to [battery=$B_{AUX}$](\Batx,\bottomLine)
            -- (\BidirectionalRightx,\bottomLine);
        \draw[dashed]
            (\BidirectionalLeftx-\border,\topLine+\border) rectangle (\BidirectionalRightx+\border,\bottomLine-\border);
        \node[align=center] at
            ({(\BidirectionalLeftx+\BidirectionalRightx)/2}, {(\topLine+\bottomLine)/2}) {(3)\\Bidirectional\\converter};
    \end{circuitikz}
    }
\end{center}
    \caption{The hardware system architecture consists of three main components: (1) an input filter to buffer the power grid against high-frequency power fluctuations,, (2) a DC-DC converter to maintain constant rack voltage, and (3) an auxiliary battery system to store or dispatch energy during transients. This configuration allows the power grid to gradually transition between different load conditions while the rack sees immediate power availability.}
    \label{fig:circuit-schematic}
\end{figure*}

Figure~\ref{fig:circuit-schematic} shows a simplified circuit schematic for the
hardware system. \name~has three hardware elements: an input filter,
a rack voltage regulator, and an auxiliary energy storage system. These 
components work in concert to remove transients from a rack's power draw
and ensure it meets grid specifications. A passive LC stage attenuates fast transients directly 
in the electrical path, while a local battery compensates for slower variations.
Assuming the rack is provided with adequate energy storage
capacity and a proper input filter, this design can be adapted to meet
\textit{any} grid specification while the rack sees immediate power
availability.

\subsection{Input Filter}
The filter shown in Figure~\ref{fig:circuit-schematic} is a second-order
passive filter with a resistive damping leg. Capacitor $C_F$ and inductor $L_F$ together store a small amount of energy that is naturally dispatched to filter out frequency content above the filter's resonant frequency, $f_f$, determined by the part values used:
\begin{equation}
\label{eq:filter_cutoff_freq_fF}
f_f = \frac{1}{2\pi\sqrt{L_F C_F}}
\end{equation}
The damping circuit composed of $R_{Da}$ and $L_{Da}$ supresses resonance between the main inductor and capacitor during transients. Because the energy density of inductors and capacitors is low, the input filter does not store very much energy---this sort of passive filtering is only viable for smoothing very short ($\leq$50 ms) transients and for filtering out noise introduced by switching components in the voltage regulator and bidirectional converter subsystems.

\subsection{DC-DC Converter}
The DC-DC converter maintains a constant voltage, $V_{OUT}$, to power the rack. The controls for the voltage
regulator in our prototype were implemented fully in hardware, meaning there is no processing time delay. This subsystem has a negligible effect on transient smoothing, but is necessary to provide immediate power availability to the rack.

\subsection{Auxiliary Energy Storage System}
Because a passive filter can't reasonably store very much energy in a confined volume, \name uses an actively controlled energy 
storage system to buffer against changes in rack power over 
longer timescales. When the rack's power drops, the storage 
system charges, absorbing the extra power from the grid. When 
the rack's power rises, the storage system discharges, 
temporarily providing power to the rack until the grid supply 
can catch up.
High-bandwidth sensors detect changes in rack power
and automatically trigger compensating current $i_B$ to/from the rack bus according to the control equation
\begin{equation}
    \label{eq:active_controller}
    \frac{d}{dt}i_B + \beta i_B + \frac{d}{dt}i_R = 0
\end{equation}
where $\beta$ is the grid-imposed limit on ramp rate from Equation~\ref{eq:ramp_rate_spec}, expressed as a fraction of the rack's maximum power utilization per second. Our prototype, for 
example, assumes $\beta=0.1$ p.u./sec, using a battery to taper fluctuating rack power such that the grid has about 30 seconds to transition between stead-state power levels.

\begin{figure}
    \centering
    \includegraphics[
        width=0.9\columnwidth,
        trim=0.5cm 0.25cm 0.3cm 0.2cm,
        clip
    ]{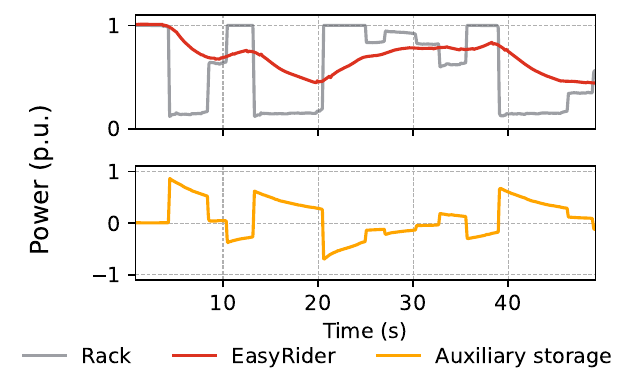}
    \caption{Smoothing shown in an \name prototype test. Grid power (red) remains smooth even though rack power (grey) fluctuates. The auxiliary energy system (orange) absorbs the difference.}
    \label{fig:simple-trace}
\end{figure}

Figure \ref{fig:simple-trace} 
demonstrates the smoothing effect of the battery 
system during a preliminary test of the \name 
prototype.

Traditional grid batteries are designed to provide power for hours, while \name only needs a few minutes worth of capacity. In our prototype, we used high-power lithium iron phosphate (LiFePO$_4$) batteries as energy storage, due to their high power-to-capacity ratio and low cost per unit energy. Supercapacitors or a combination of different energy storage technologies could also meet \name's storage needs. The key sizing requirements for this system are defined in Appendix~\ref{app:component-sizing}.

\subsection{Filter Response}
\name's hardware consists of two filters (the passive input filter
and the controlled energy storage system), and its net response to transients 
is the simple multiplication of their individual responses.  Careful component selection can provide adequate filtering to keep the rack in compliance with grid-imposed frequency content specifications as in equation~\ref{eq:frequency_content_spec} in Section 3. Appendix~\ref{app:component-sizing} describes
how the correct sizes are derived from the rack power rating and grid specifications.

\begin{figure}
    \centering
    \includegraphics[
            width=0.9\linewidth,
            trim=0.0cm 0.25cm 0.0cm 0.25cm,
            clip
        ]{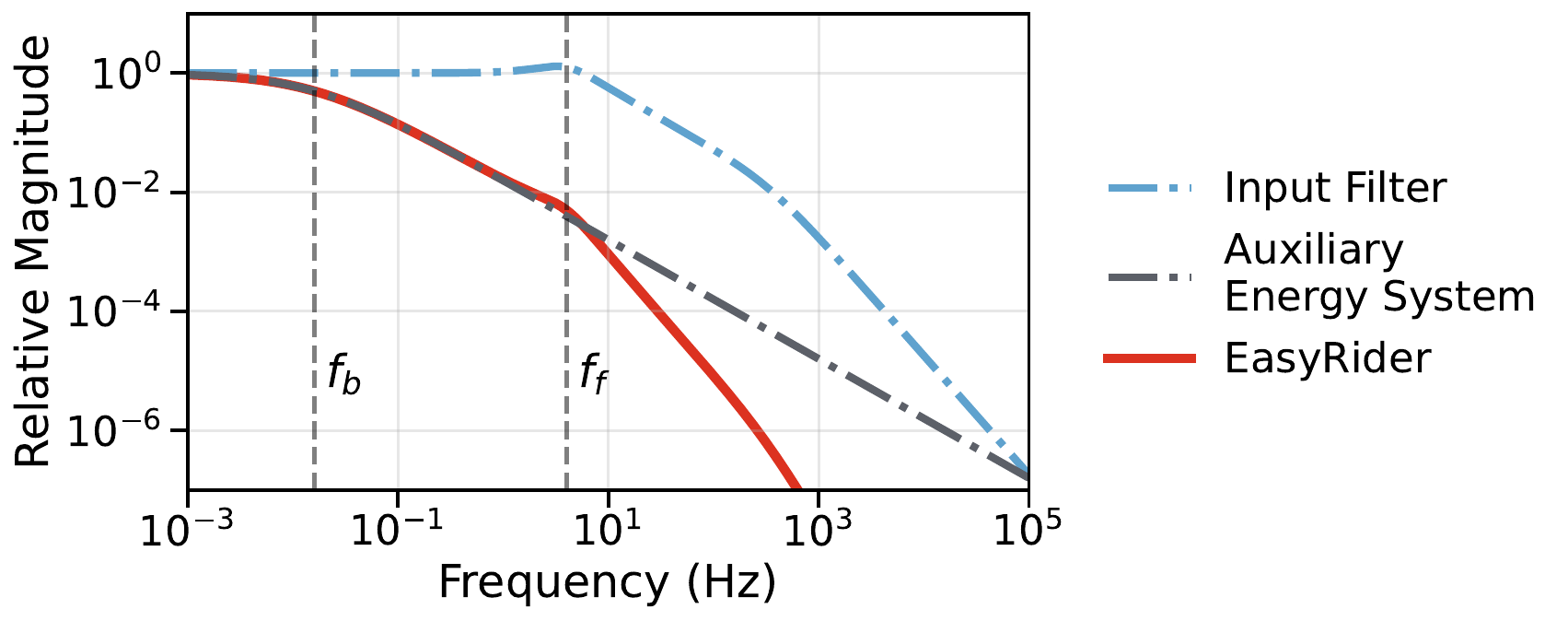}
    \caption{Our~\name~prototype's frequency response, showing the combined effect of the input filter and controlled energy storage system. The input filter attenuates fluctuations above $f_f$, while the auxiliary energy compensates for fluctuations above $f_b$. Together, they ensure the rack meets grid specifications. ``Relative Magnitude'' indicates the magnitude of fluctuations seen by the DC distribution grid relative to those drawn by the rack, $\left| \frac{\tilde{i}_R}{\tilde{i}_{DC}} \right|$.}
    \label{fig:frequency_response}
\end{figure}

The passive filter pictured in Figure~\ref{fig:circuit-schematic} attenuates frequencies by up to an additional 99\% for every 10x increase in frequency above the cutoff frequency $f_f$ (calculated in equation~\ref{eq:filter_cutoff_freq_fF}). Our~\name~prototype used a cutoff 
frequency $f_f\approx 4$~Hz, and Figure~\ref{fig:frequency_response} shows how the filter attenuates fluctuations above this frequency. Without the controlled auxiliary energy system, a sinusoidal change in rack power with
$f=1$~Hz will not be dampened at all by the input filter, while a fluctuation
at $f=1000$~Hz will be cut by a factor of $\approx$ 1000, 
as observed by the grid.

The auxiliary energy system is also a filter. Because this system is actively controlled, its cutoff frequency follows from equation~\ref{eq:active_controller} such that
\begin{equation}
\label{eq:fb}
    f_b=\frac{\beta}{2\pi}
\end{equation}
The net filter response of this system is that rack frequency content is attenuated by a factor of 10 for every 10x in frequency above $f_b$.

These two filters compound. Figure \ref{fig:frequency_response} shows the total
frequency response of the \name system as the product of the input filter and auxiliary
system's responses. 



\section{Battery Lifetime Management}
\label{sec:software-control}
 
\name's hardware path handles transients autonomously, with no software in the loop.
But because the battery charge and discharge efficiencies ($\eta_c$ and $\eta_d$ respectively) are not perfect, an additional controller is required to manage the battery's SoC over longer timescales.

Every charge-discharge cycle incurs round-trip losses on the order of
$1 - \eta_c \eta_d$ of the energy exchanged, and these losses accumulate
over hours of training into a monotonic SoC drift. This can be either upward when set-point
bias dominates, or downward when resistive losses dominate.  
Left uncorrected, the battery eventually
saturates against its upper or lower safe bound, losing the symmetric
headroom it needs to smooth the next transient. Dwelling at an extreme
SoC also accelerates idle-time aging~\cite{Gantenbein2019CapacityFade,Takei2001CycleLife,Soto2022MicroCycles,calendar-aging}.

The software controller's purpose is to counteract this drift.  It
periodically reads the battery's state of charge from the battery
management system and issues milliamp-scale corrective currents through the bidirectional converter.  Because the corrective current is orders of
magnitude below the rack's transient current, the controller does not
interfere with the hardware's filtering.
We decompose the controller into two loops that operate on different
timescales: an outer loop that selects the SoC target and an inner loop that drives the battery toward the target.

\noindent\textbf{Outer Loop:} A slow \emph{outer loop}, updated on regime changes and
refreshed every few minutes, selects target SoC based on reducing battery aging~\cite{battery_aging_model}.  
During active training, the target is a mid-band
value~$S_{\mathrm{mid}}$ chosen to maximize symmetric charge and
discharge headroom.  The system will also detects when the battery is sitting idle (i.e. during maintainence windows) and will automatically lower the target SoC in order to reduce voltage-dependent 
idle-time aging~\cite{battery_aging_model,calendar-aging}.
 
\noindent\textbf{Inner Loop:} A faster \emph{inner loop}, executed every 5\,s, drives the
battery toward the current target by solving a small convex program over
a receding horizon of~$H$ intervals. The
controller applies only the first action from each solve and re-optimizes
at the next interval with a fresh SoC reading from the BMS. The controller is designed to avoid unnecessary cycling and to act gradually enough that it doesn't conflict with the load-smoothing function of the broader system. 
 
The controller depends on three groups of
parameters: (1) battery properties such as max charge/discharge current and round trip efficiency, (2) outer-loop policy such as the mid-band SoC and idle-time SoC, and (3) inner-loop weights such as the tracking error, maintenance-current magnitude, and command smoothness. These are all set
once at deployment from the battery datasheet and the desired correction
timescale, with no per-workload tuning. The full formulation, including the storage-target computation, the QP
objective and constraints, and the normalization of tuning
weights appears in Appendix~\ref{app:controller-formulation}.
\section{Evaluation}
In this section, we experimentally demonstrate~\name~'s ability to keep training loads grid-compliant without affecting jobs. We also compare it to other solutions and analyze the overheads and system lifetime tradeoffs.

\begin{figure}
    \centering
    \includegraphics[width=0.6\linewidth]{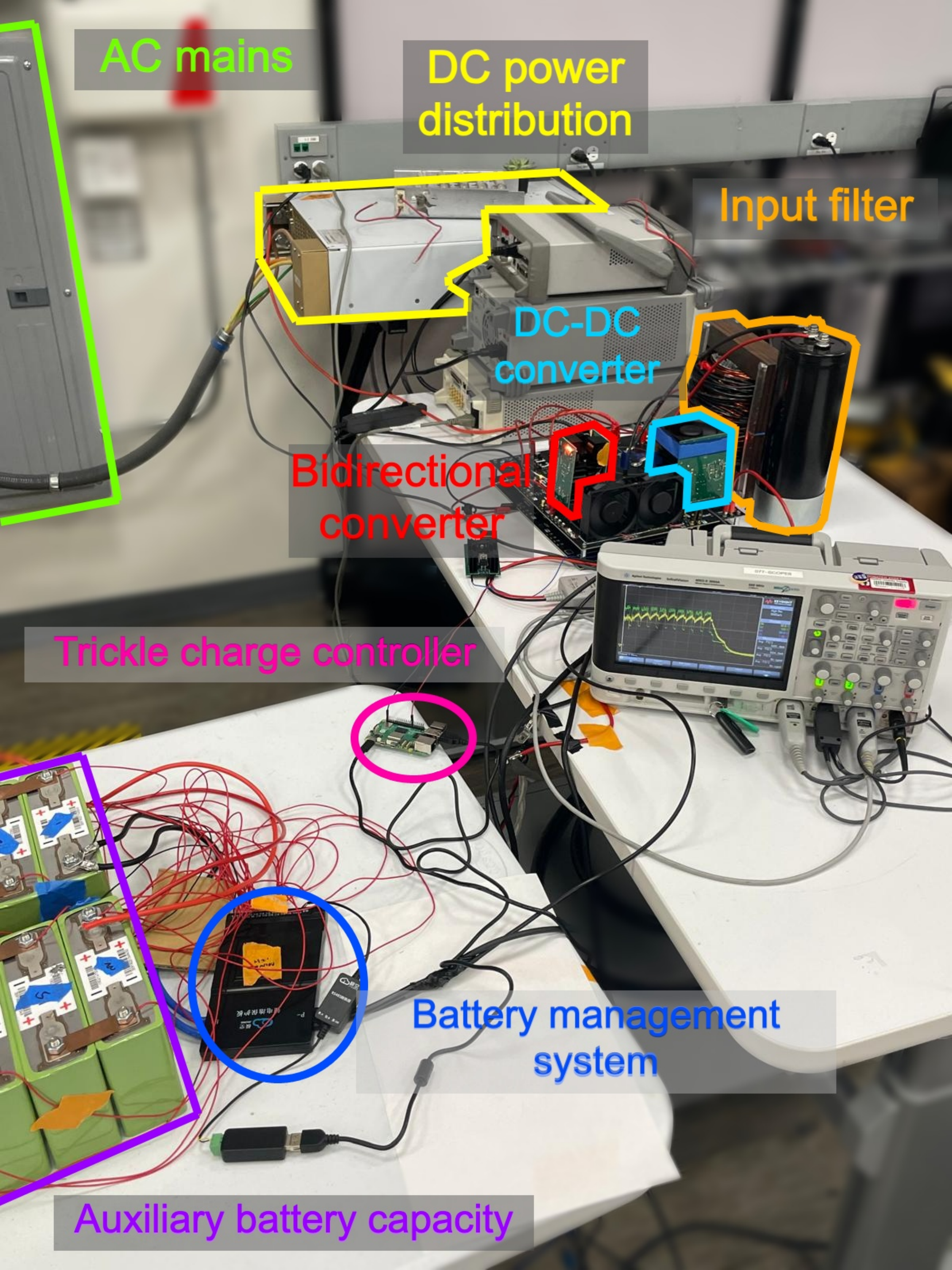}
    \caption{Photo of the built \name~prototype system.}
    \label{fig:system-photo}
\end{figure}

\subsection{Experimental Setup}
\noindent\textbf{Prototype rating:}
To evaluate the~\name architecture, we constructed a prototype of the system design outlined in Sections~\ref{sec:hardware-design} and~\ref{sec:software-control}.
This prototype, pictured in Figure~\ref{fig:system-photo}, is rated to deliver 10kW of power to a 400 $V_{DC}$ load.
It is equipped with a 74Ah battery bank with a max discharge rate of 2.4C. It should be noted that this prototype is intended solely as a proof of concept, and was not optimized for minimum volume. The 10 kW rating should be taken as a reflection of a constrained research budget rather than any sort of bound in the design space.\footnote{It is standard to use power supplies in parallel to achieve higher power ratings in rack-scale power systems. For example, the existing Mt. Diablo 1 MW rack designs combine 11 18 kW PSUs into a 180 kW, 1U shelf\cite{mtdiablo}.}

\vspace{1ex}
\noindent\textbf{Workloads and traces:}
Cluster-scale power traces of frontier-model training jobs are not generally made public, but we were able to use a normalized trace from~\cite{microsoft_power_stabilization} in addition to power profiles we measured from training a
125M parameter GPT-style LLM on a decommissioned 2-GPU NVIDIA Titan-X server 
blade from our lab.

\subsection{Ramp Rate \& Frequency Content Compliance without Training Changes}
\noindent\textbf{Benchmark specifications:}
As discussed in Section~\ref{sec:problem-statement}, the 
maximum allowable ramp rate $\beta$ and parameters 
$\alpha$ and $f_c$ defining restrictions on the frequency 
content of the grid power trace will be set by local
grid operators' requirements. To demonstrate the 
smoothing effect, we designed our \name~prototype under the assumption that the datacenter is allowed to ramp at 
a maximum $\beta=0.1$ (10\% of rated power per second) 
and that the grid imposes a limit $S(f) < \alpha = 10^{-4}
$ on the normalized magnitude of frequencies $f$ above 
$f_c = 2~\text{Hz}$. This spec is in line with prior work~\cite
{microsoft_power_stabilization,semianalysis2025,li2024unseenaidisruptionspower}
and addresses the band of frequecies from 0.1-10 Hz that 
can damage generators and turbines as noted by NERC~\cite
{nerc2019oscillation,NERC2021_oscillations, nerc2025largeloads}. 

\begin{figure}
    \centering
    \begin{subfigure}[b]{0.9\columnwidth}
        \includegraphics[
        width=0.95\linewidth,
        trim=0.1cm 0.35cm 0.10cm 0.25cm,
        clip
    ]{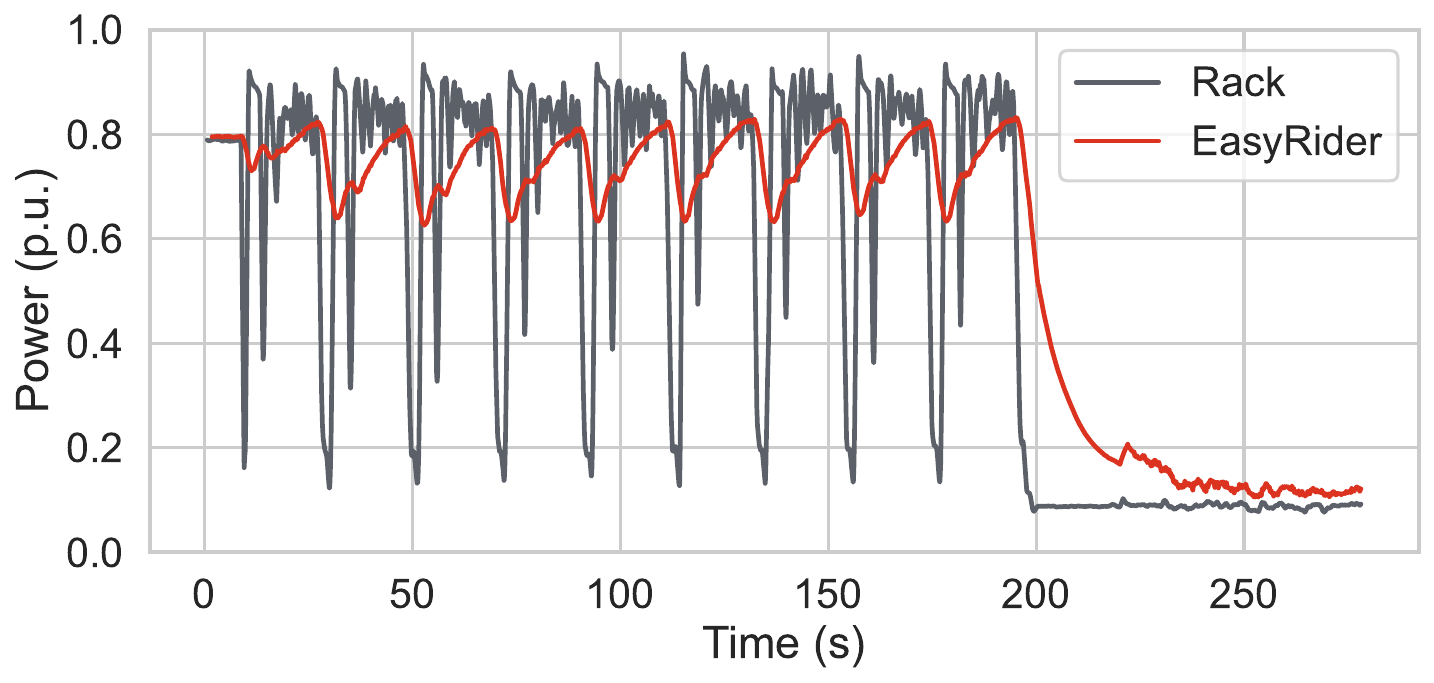}
        \caption{Power trace.}
        \label{fig:easy-rider-msft-trace-192v}
    \end{subfigure}
    \hfill
    \begin{subfigure}[b]{0.9\columnwidth}
        \includegraphics[
        width=0.95\linewidth,
        trim=0.1cm 0.35cm 0.1cm 0.25cm,
        clip
    ]{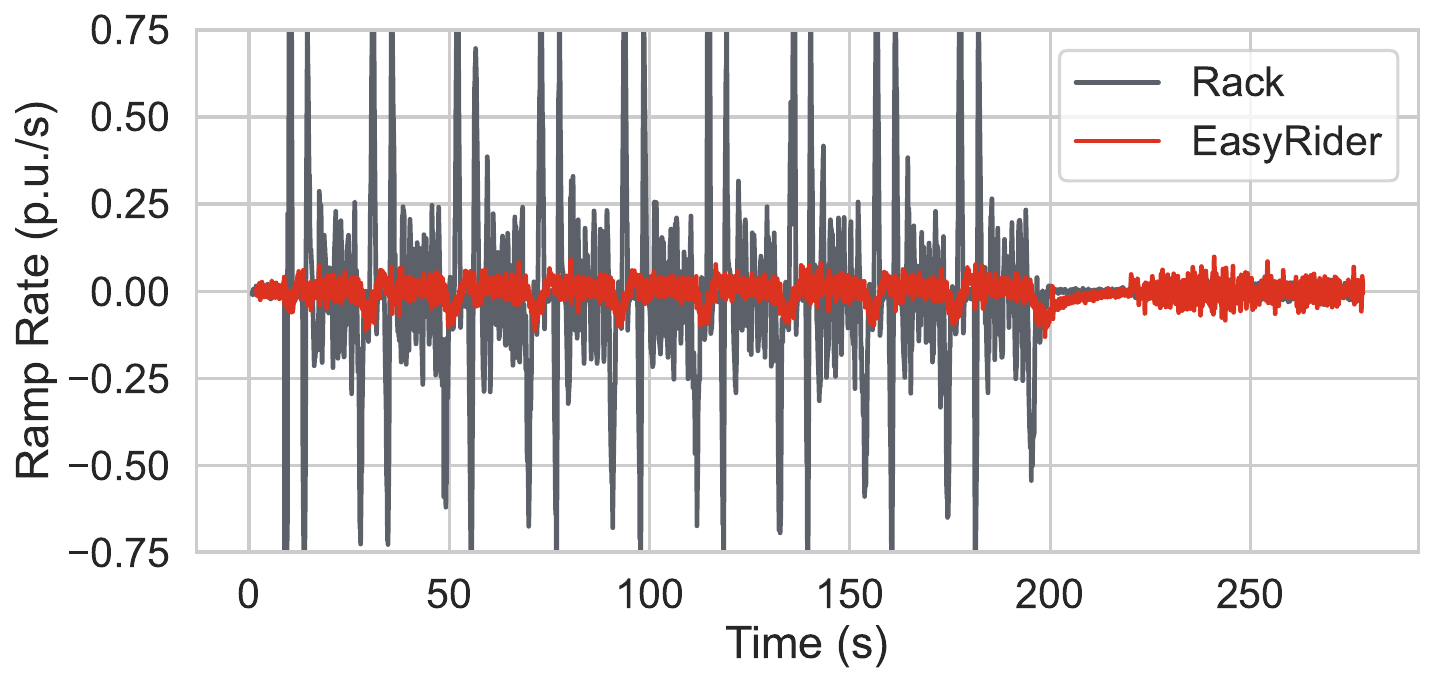}
        \caption{Ramp rate.}
        \label{fig:ramp-rate-msft-trace-192v}
    \end{subfigure}
    \caption{(a) Conditioned power trace using \name to power a DC load with a jittery training power trace. (b) Corresponding ramp rate of power drawn from the grid compared to the unconditioned ramp rate as a function of time. The \name~prototype is able to constrain the rack's ramp rate to less than $\pm10\%$ of its rated power per second.}
    \label{fig:ramp-rate-compliance}
\end{figure}

\noindent\textbf{Ramp rate compliance:}
Figure \ref{fig:easy-rider-msft-trace-192v} shows the effect of our~\name prototype smoothing the last few minutes of a training run. While the rack power trace exhibits sharp power swings at each communication event and an abrupt drop at job termination, the \name-conditioned trace transitions gradually and exhibits a lower peak power draw. Figure~\ref{fig:ramp-rate-msft-trace-192v} shows the ramp rate across the same time period, demonstrating that \name~successfully smooths the rack power trace to ensure that the ramp rate never exceeds 10\% of the rack's rated power per second.

This behavior is independent of the training job's power 
profile, therefore complying with the grid ramp-rate 
specification without modifying the workload. This decouples
grid compliance from job scheduling, as any training workload can run unmodified. Importantly, the guarantee composes across racks and rows---because each
\name-equipped rack presents a power waveform with $\mid dP/dt\mid \le \beta$,
the aggregate datacenter ramp rate is likewise constrained.\footnote{Appendix~\ref{app:smoothing-at-scale} provides additional explanation on smoothing effects at a cluster scale.} This allows operators
to reason about campus-wide limits in terms of per-rack design rather than
per-job coordination.


\begin{figure}
    \includegraphics[
        width=0.95\linewidth,
        trim=0.0 0.35cm 0.0cm 0.3cm,
        clip
    ]{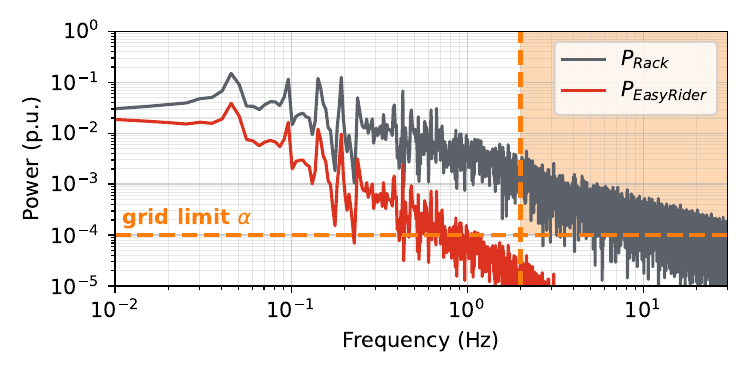}
    \caption{The filtering effect of \name~keeps harmonic content below
    a grid-imposed limit $\alpha$ for frequencies above $f_c = 2~\text{Hz}$, even though
    the rack power trace contains significant energy in this band.}
    \label{fig:system-frequency-response}
\end{figure}

\noindent\textbf{Frequency content compliance:}
Figure~\ref{fig:system-frequency-response} shows the rack and \name power
traces from Figure~\ref{fig:easy-rider-msft-trace-192v} broken into their respective frequency components. The combined effect of the input filter and battery system is enough
to shift the entire \name~power spectrum out of restricted zone.

Because \name~implements a fixed
transfer function (Figure~\ref{fig:frequency_response}) at the rack-to-power-distribution connection, 
any training job whose raw power spectrum falls at or below the grey
curve will, after conditioning, satisfy the same grid constraint without needing to change the model,
scheduler, or GPU firmware. As with ramp rate compliance, the guarantee composes across racks and rows: each
\name-equipped rack enforces the same per-rack bound on $S_{\mathit{grid}}(f)$, so a hall of 
racks behaves like a collection of ``tamed'' loads that can be integrated under a
campus-level interconnection agreement. \name~turns arbitrary high-frequency power
fluctuations from training into a waveform whose worst-case harmonic content is known and
bounded by design. 

\subsection{Energy Efficiency Against Software Burn-Based Solutions}
Other approaches to enforcing ramp-rate limits, as discussed in
Section~\ref{sec:background:approaches}, either rely on cluster-wide
coordination or use proprietary hardware that we cannot reproduce. The most
directly comparable, software-only mechanism is to inject ``burn'' kernels that
artificially raise GPU utilization to a target power level. We therefore
compare \name~to a software burn-based solution on our decommissioned
2-GPU Titan~X blade.

To implement the software burn, we profile GEMM kernels to derive a mapping
between duty cycle and GPU power, then use this mapping to schedule additional
matrix multiplications that maintain or ramp to a desired power setpoint. Full
details of this implementation appear in Appendix~\ref{app:software-burn}.
Figure~\ref{fig:titanx-power-comparison} shows the resulting normalized power
traces for the raw Titan~X workload, \name, and software burn. We delay the
start of the Titan~X trace by approximately 41\,s to account for the warm-up
period required by software burn, and normalize all traces to the Titan~X
blade’s TDP.

Not from Figure~\ref{fig:titanx-power-comparison} that~\name~has a lower peak power than either the GPU burn or bare rack traces. In this test, \name~used 19\% less energy than the software burn, largely because it does not
require any additional warm-up period or changes to the training code.
While \name~does incur some losses in its battery and power electronics, these manifest as a small additional
energy sink over weeks of training, whereas software burns waste energy throughout every second of the job.

\begin{figure}
\includegraphics[width=\linewidth]{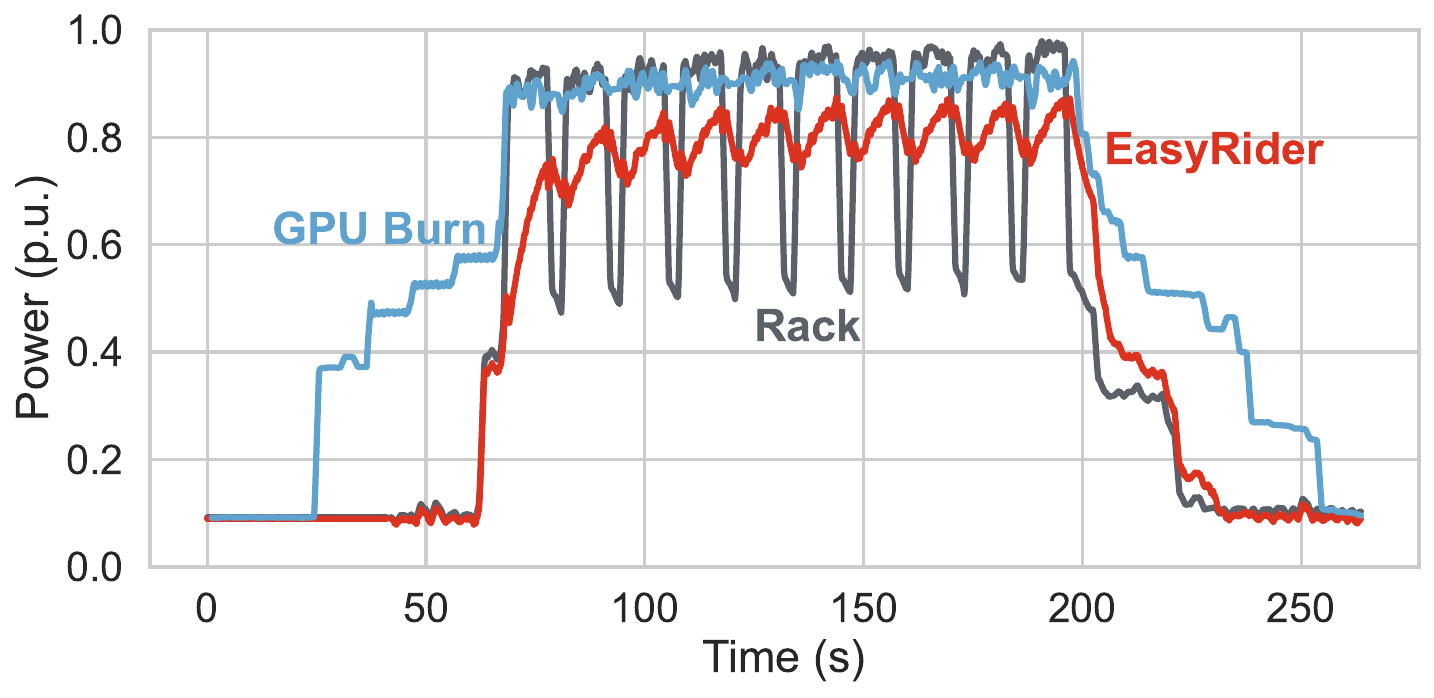}
\caption{Normalized power of the \name~prototype and a GPU burn smoothing a Titan~X trace.}
\label{fig:titanx-power-comparison}
\end{figure}

\subsection{Energy Storage Stability}

\begin{figure}
    \centering
    \includegraphics[
        width=0.95\linewidth,
        trim=0.0 0.25cm 0.0cm 0.25cm,
        clip
    ]{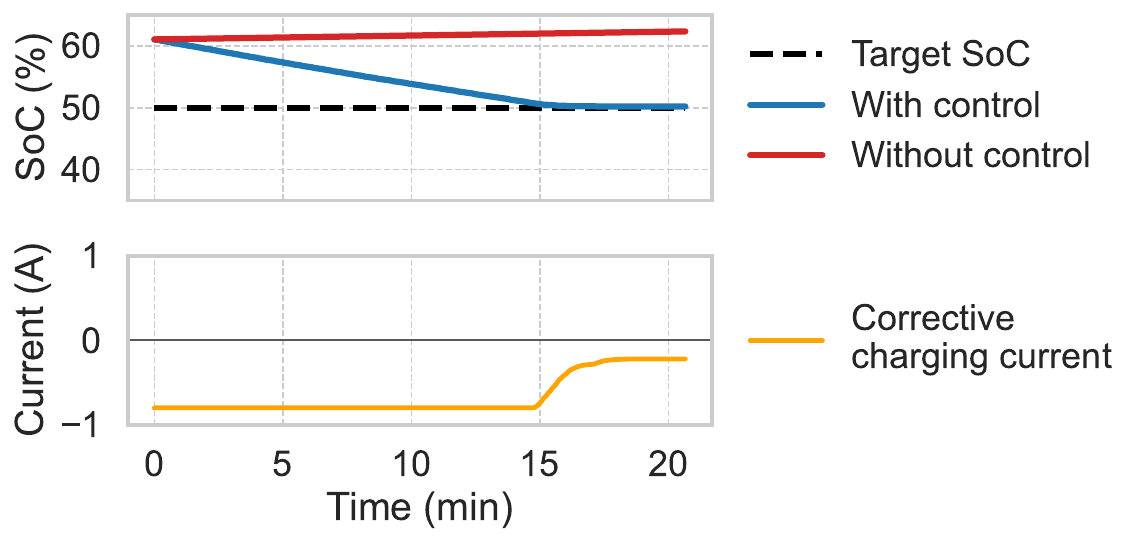}
    \caption{Battery adjustment for an SoC that is over the desired setpoint. 
    Our control system updates the corrective current every 5 seconds to return to $S_{\mathrm{mid}} = 0.5$. 
    Without this correction, the battery would drift slowly towards the upper bound.}
    \label{fig:battery-corrective-soc-overtarget}
\end{figure}

As discussed in Section~\ref{sec:software-control}, our software controller exists to counteract gradual SoC drift without interfering with the hardware's filtering.

Figure~\ref{fig:battery-corrective-soc-overtarget} demonstrates this mechanism in practice. 
After a few hours of operation without software control, our system drifts to approximately 62\% SoC. 
As soon as we begin this experiment, our software controller allows the inner-loop QP (Appendix~\ref{app:controller-formulation}) 
to issue corrective currents while the rack runs the training trace.  
The ``with software'' trace converges to $S_{\mathrm{mid}}$ within approximately 20 minutes. 
The ``without software'' trace shows the SoC without corrective current, gradually drifting as set-point bias in the hardware system pushes the SoC toward the upper safe bound.


This experiment validates EasyRider's SoC regulation capability.
The software stack adds no per-workload tuning. Since $S_{\mathrm{mid}}$ and the QP weights are set once from the battery datasheet, the controller keeps the battery in its optimal operating band across arbitrary training traces.

\subsection{Projected System Lifetime}
The only components in this system that are likely to degrade significantly with time are the battery cells. Because that degradation builds up over years, we estimate the pack's lifetime with a semi-empirical lithium-iron-phosphate aging model~\cite{Suri2016ControlOriented}. We drive the model with the actual current our controller produces while smoothing a real training trace, holding the pack near mid-charge. On a measured H100 trace the cells cycle at only about 0.05C, and the model predicts roughly 60 years before capacity falls to 80\%. A trace with deeper, more frequent power swings raises the average rate toward 0.26C and shortens the life to about a decade. Even the shorter figure meets the long service life for which lithium-iron-phosphate cells are known~\cite{Yang2025LFPAging}, so across these workloads \name's cycling is not what limits the battery's life.

This may seem surprising given that the batteries are utilized almost constantly while the system is active---however, because the battery pack size is limited by the peak system power rather than the stored energy, the battery in our prototype uses less than 1\% of its capacity. ``Micro-cycling'' at a mid-range SoC has been shown to extend the total lifetime charge throughput for a battery far beyond what would be possible were the battery constantly cycled with 100\% depth-of-discharge~\cite{Guena2006DepthDischarge,Soto2022MicroCycles,Ceraolo2016MicroCycles,Ovejas2019Hysteresis}. Although buffering rack power fluctuations diverts some mid-to-high-frequency (>100 Hz) power to the battery, these components are dwarfed by the low-frequency content (<1 Hz) in terms of total charge movement and consequently have little effect on the battery lifetime~\cite{Amamra2020ElectricVehicle}. Assuming the cells are kept reasonably cool, higher-frequency (>1 kHz) cycling will not degrade the batteries because the cells' double-layer capacitance between the electrodes and electrolyte is charged/discharged rather than cycling actual lithium ions~\cite{Jossen2006,Uno2011HighFrequency}.
\section{Discussion}

\noindent\textbf{Costs/savings:}
Our 10 kW prototype had a bill of materials cost of approximately \$3,500USD. A large fraction of this cost comes from the 74 Ah battery pack, which is intentionally oversized relative to the requirements derived in Appendix~\ref{app:component-sizing}. We expect that a production design would benefit from economies of scale and further design optimizations that would bring capital expenses down even further. However, in our current deployment we achieve \$0.35/W, which, scaled to the power level of a GB200 rack, is less than 1.25\% of the estimated \$3.7 million USD rack cost.

Compared to any of the methods utilizing power burns to keep GPU utilization high, \name~offers significant energy cost reductions. A 30 MW cluster draws the equivalent power of $\approx$2400 average U.S. homes~\cite{eia_residential_electricity_use}. At such a power level, eliminating even 5\% of wasted electricity for the same cluster at a conservative 10\textcent/kWh leads to over \$1.3 million USD in savings every year.

\noindent\textbf{Rack-level placement:}
\name is designed for deployment at the rack or row level, rather than higher up in the energy distribution heirarchy, so its deployment can easily and incrementally match datacenter needs. Deploying it alongside a centralized facility-level BESS requires knowing the peak facility ML load at the time of installation. A rack-level application scales better because installed battery capacity will match the installed load. Because trainign presents a synchronous load across racks, there is a far less inherent smoothing in the load across the facility than in traditional cloud computing datacenters: incremental deployment therefore does not lead to overprovisioning.

\noindent\textbf{Fault tolerance and complexity:}
Reliability is critical for any power system. EasyRider is composed of the same general components as any modern datacenter power hierarchy: power supplies already contain DC-DC regulators and passive input filters, and UPS systems already use batteries for backup energy. EasyRider does not introduce new battery management, redundancy, or fault-recovery mechanisms; its failure assumptions are the same as existing power infrastructure. In modern datacenter architectures, failures at the PDU and PSU level are a third order concern due to the tremendous reliability of these components---they have been engineered reliably for over 100 years. Furthermore, if an~\name unit were to fail prematurely, the per-rack solution is an advantage as the worst that happens is a single rack's transients go unfiltered.

\noindent\textbf{Minimal and isolated software.}
The software stack is deliberately small and decoupled from training jobs. Its
only roles are to measure battery state-of-charge and current and to issue
slow, corrective current adjustments that keep the battery in a healthy SoC and
voltage range. This
implementation is identical across racks and does not interact with model code,
frameworks, or schedulers.

\section{Related Work}

\noindent\textbf{Characterization of AI Training Power Dynamics.} 
Large swings in compute load have been documented in HPC systems for over a decade~\cite{bates2015,shin2021revealing,patki2025global,stewart2019grid}. 
This literature studies how large compute clusters interact with the grid, 
quantifies problematic ramp rates and oscillations, 
and considers scheduler-level mitigation. 
Our setting is different, since recent work has shown that large AI training jobs 
create tightly synchronized, multi-megawatt swings with distinct temporal 
structure~\cite{li2025ailoaddynamicsapower,li2024unseenaidisruptionspower,semianalysis2025}. 
\name builds on that observation, but targets mitigating these 
transient risks, not just characterizing the loads.

\noindent\textbf{Datacenter Power Management.}
Prior datacenter power-management systems treat power as a shared resource to allocate, cap, or oversubscribe through 
cluster-level control~\cite{wu2016dynamo,li2019capmaestro,thunderbolt,argo2016ellsworth,zhang2021flex,hsu2018smoothoperator,piga2024dvfs,kumbhare2021prediction}. 
Similar work uses batteries and UPSes for peak shaving, power capping, or carbon shifting~\cite{zheng2014teshave,deepPM,bianchini2024powermanagement,Govindan2011EBuff,Acun2023CarbonExplorer}. 
These systems determine \emph{when}, \emph{where}, and \emph{how much} power workloads may consume. 
\name, in contrast, formulates transient mitigation as a frequency-domain filtering problem, regulating how quickly the power draw can change. It is therefore complementary to existing control-plane mechanisms, but provides an altogether different service.

\noindent\textbf{AI/ML Training Power Management.} 
Recent work on AI training power and energy management 
reduces energy consumption by changing training behavior through profiling, scheduling, DVFS, or 
power capping~\cite{chung2024reducing,zeus,koszczal2023performance,zhao2023sustainsupercomputing,wang2022dynamicgpuenergyoptimization,choi2023envpipe}. 
The goal in this literature is to improve job-level energy efficiency or fit workloads within cluster power budgets. 
\name does not modify the training job; instead, it leaves the workload unchanged and reshapes the resulting power draw in the rack PDU.

\noindent\textbf{Industry characterization and proposals.}
Existing industry proposals mitigate training transients at different points 
in the stack: some shape computation through burn or power 
capping~\cite{semianalysis2025,nvidia-caps}, 
while others buffer power at the platform or site 
boundary~\cite{xai-power-stabilization}. 
Related analyses of AI load dynamics have also 
clarified the grid-side risk created by synchronized 
training loads~\cite{li2024unseenaidisruptionspower,semianalysis2025}. 
Other proposals consider rack-level storage coordinated with 
software control~\cite{microsoft_power_stabilization}. 
Across these efforts, the hardware and software roles 
in transient mitigation remain only loosely separated. 
\name is able to manage transients in hardware which is fast
and path independent from the software stack.
\section{Conclusion}
In this paper we presented \name, a per-rack power system that
conditions training power before it reaches upstream distribution and the grid.
\name combines a passive filter, an actively controlled 
rack-scale battery, and a DC regulator, powering the rack while also enforcing grid-facing limits on ramp rate and frequency 
content, all without modifying the training stack. A 
lightweight optimization controller monitors and maintains battery health over time.
Using a 10\,kW prototype on
normalized cluster traces and real GPU training workloads, we show that \name~meets these grid constraints while incurring substantially lower energy and
runtime overheads than software burn-based approaches. Because it sits entirely
behind the rack PDU and relies only on local sensing and control, \name~can be
incrementally deployed across existing and future high-voltage DC racks,
providing a practical path to grid-safe AI clusters as model and rack power
continue to scale.

\bibliographystyle{ACM-Reference-Format.bst}
\bibliography{references.bib}
\appendix
\clearpage 
\begin{figure*}[h]
  \centering
  \includegraphics[width=0.9\textwidth]{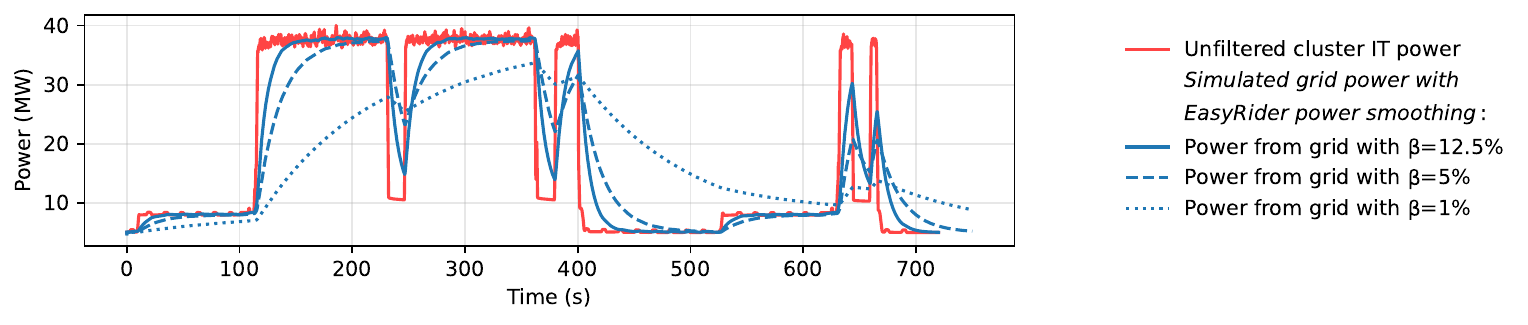}
  \caption{Expected smoothing behavior of a 40 MW training cluster where every rack is equipped with an \name power supply, vs. the unfiltered case.
  $\beta$ represents the \name-enforced maximum rack power ramp rate (see Section \ref{sec:problem-statement}), as a proportion of maximum rated rack power per second.
  The IT trace (red) is scaled from actual measurements from running a training job on H100 GPUs.
  The highest ramp rate recorded in this trace occured when the system experienced a computation fault observed around 400 s, causing a near-instantaneous drop in power. At this point, the red trace falls at a rate of 193.7 MW/sec (11.6 GW/min), which is far outside the range of what conventional generators could compensate for. Also note that such a computation fault would be difficult to predict in order to smooth using a scheduled power burn, but the plot shows that \name still provides smoothing because it does not depend on software telemetry to detect power fluctuations.}
  \label{fig:power-at-scale}
\end{figure*}

\section{Explanation of Smoothing at Scale}
\label{app:smoothing-at-scale}
The smoothing effect of \name~at the campus-wide scale follows from the fact that 
the total datacenter power use is a sum of all the individual system demands. For a cluster spread across N racks, we can break
down the total cluster IT load $P_{IT}(t)$ into the sum of the instantaneous power demands of each rack:
\begin{equation}
  \label{eq:power-sum}
  P_{IT}(t) = \sum_{i=1}^N P_i(t)
\end{equation}
In syncronous training, because all the individual power traces across a cluster are essentially the same,
\begin{equation}
  \label{eq:cluster-power}
  P_{IT}(t) = N \cdot P_i(t)
\end{equation}
and furthermore because the DFT is a linear function, it also allows scaling by a linear multiplier---the power spectrum for the cluster is proportional to the power spectrum of the individual racks operating in synchrony:
\begin{equation}
  \label{eq:cluster-spectrum}
  S_{IT}(f) = N \cdot S_i(f)
\end{equation}
Although the prototype demonstrated in this paper is only rated for 10 kW as a proof of concept and does not singlehandedly handle enough power to smooth grid-scale fluctuations, the scaling relations in equations \ref{eq:cluster-power} and \ref{eq:cluster-spectrum} assert that the normalized results we show in the paper for a single rack would look identical at the cluster scale, were every rack equipped with an \name power supply.

Figure \ref{fig:power-at-scale} shows the expected smoothing behavior on a 40 MW training cluster were every rack equipped with an individual \name power supply. Because \name~smooths each rack's power, the aggregate cluster power $P_{IT}$ is also smoothed.

\section{Hardware Components: Values and Sizing}
\label{app:component-sizing}

\noindent{\bf Energy storage capacity:} Suppose we are using \name
to ride through the power transients of a rack with a themal design power (TDP) of
$P_{RATED}$. The design depends on an adequately sized energy storage system,
whether using batteries, supercapacitors, or any other storage mechanism. 
The DC-DC regulator stage maintains the voltage at the input of the rack at a constant
$V_{OUT} = V_{DC}$, so power diverted to the auxiliary energy storage branch at any given
time $(t)$ is
\begin{equation}
\label{eq:ESS_power_instantaneous}
P_B(t) = V_{DC} \cdot i_B(t)
\end{equation}

\name`s energy storage system is controlled in our design such that the current
$i_B$ is fixed by the differential equation
\begin{equation}
\label{eq:ESS_control}
\frac{d}{dt}i_B + \beta\cdot i_B + \frac{d}{dt}i_R = 0
\end{equation}
which ensures that the maximum ramp rate that the \name system imposes on the
grid can never exceed $\beta\cdot P_{RATED}$, even if the rack were to turn off
altogether. ($\beta$ is chosen to meet the system ramp rate restriction as shown in Figure \ref{fig:microsoft_trace-time}, discussed in Section \ref{sec:problem-statement}.)

If we assume that at time $t=0$, $i_R$ has been constant at some current $I_1$ for some time, and then over a period of some time it transitions to some current $I_2$ and holds steady, the net energy (in joules) stored in the during the transient under ideal conditions is
\begin{equation}
\label{eq:ESS_energy_general}
\Delta E_B = V_{DC}\int_{0}^{\infty} i_Bdt = -\frac{V_{DC}}{\beta}\int_{0}^{\infty}\frac{d}{dt}(i_R+i_B)dt
\end{equation}
From equation \ref{eq:ESS_control} we know that the battery current decays to zero when the rack current is constant. Then
\begin{equation}
\label{eq:ESS_energy_solved}
\Delta E_B = -\frac{V_{DC}}{\beta}[i_R(t)+i_B(t)]_{t=0}^{t=\infty} = \frac{V_{DC}}{\beta}(I_1-I_2)
\end{equation}
The maximum change in rack power as a proportion of total TDP is
\begin{equation}
\label{epsillon}
\epsilon = \frac{P_{RATED}-P_{MIN}}{P_{RATED}}
\end{equation}
where $P_{MIN}$ is the minimum ($\geq 0$) rack power in watts. Because the energy storage system won't ever charge unless the rack power has generally decreased, and the system won't ever discharge unless the rack power has generally increased, the maximum magnitude of $\Delta E_B$ in equation \ref{eq:ESS_energy_solved} occurs for $I_1=\frac{P_{RATED}}{V_{DC}}$, $I_2=\frac{P_{MIN}}{V_{DC}}$, the maximum and minimum possible rack currents, repectively:
\begin{equation}
\label{eq:ESS_energy_max}
\left|\Delta E_B\right| \leq \Delta \left. E_B\right|_{I_1 = \frac{P_{RATED}}{V_{DC}}, I_2 = \frac{P_{MIN}}{V_{DC}}} \end{equation}
Therefore we can conclude that the net energy stored during any rack power trace is bounded by
\begin{equation}
\label{eq:ESS_bounded}
\Delta E_B \leq \frac{\epsilon}{\beta}P_{RATED}
\end{equation}
Finally, if you are restricted to using only some proportion $\gamma$ of the total capacity of the energy storage mechanism---as in the case of batteries, which may need to be kept in a 40-60\% state of charge to prevent rapid aging---the minimum viable storage capacity $E_B$ in joules is
\begin{equation}
\label{eq:ESS_capacity}
E_B \geq \frac{\epsilon}{\gamma\beta}P_{RATED}
\end{equation}

\vspace{1ex}\noindent{\bf Energy storage power rating:} The energy storage
system must also be capable of sourcing or sinking power at a sufficient rate
to maintain compliance with grid ramp rate specifications. From equation
\ref{eq:ESS_control} we can see that the maximum power that the energy storage
system must be capable of sourcing or sinking occurs when the rack power
changes instantaneously from its maximum to minimum value or vice versa. It
follows that the energy storage system needs to be rated to charge or discharge
at a power level of at least
\begin{equation}
\label{eq:ESS_power}
P_B \geq \epsilon P_{RATED}
\end{equation}
where $\epsilon$ is as defined in equation \ref{epsillon}.

\vspace{1ex}\noindent{\bf Input filter components:} The input filter's primary
function is to attenuate high-frequency power fluctuations in order to comply
with the frequency content specification laid out in Section
\ref{sec:problem-statement}. The control dynamics of the energy storage system
shown in equation \ref{eq:ESS_control} already ensure that power fluctuations
with harmonic content above $f_b = \frac{\beta}{2\pi}$ Hz are attenuated by a
factor of 10 for every 10x increase in frequency. Because this may not be adequate on
its own, the input filter provides additional attenuation of higher-frequency
power fluctuations.

A second-order LC filter like the one shown in Figure
\ref{fig:circuit-schematic} attenuates rack power fluctuations by a factor of
as much as 100 for every 10x increase in frequency above its cutoff frequency $f_f$. Depending
on the characteristics of the rack power profile, the cutoff frequency  is
chosen such that the grid power harmonic content is acceptable under the grid
specifications. Because $f_f$ is a function of the filter component values, the
inductance $L$ and capacitance $C$ of the filter should be chosen to achieve
the desired cutoff frequency using the standard formula for a second-order LC
filter:
\begin{equation}
\label{eq:filter_LC}
f_f = \frac{1}{2\pi\sqrt{LC}}
\end{equation}
The total system frequency response is shown in
Figure~\ref{fig:frequency_response}.

\section{Controller Formulation}
\label{app:controller-formulation}
 
This appendix states the outer- and inner-loop optimization problems
solved by \name's software controller (Section~\ref{sec:software-control}).
 
\subsection{Outer Loop: SoC Target Selection}
\label{app:outer-loop}
 
The outer loop selects a target~$S^*$ from two modes:
 
\paragraph{Active mode ($S^* = S_{\mathrm{mid}}$).}
During training, the target is fixed at~$S_{\mathrm{mid}}$ to preserve
symmetric headroom.
 
\paragraph{Storage mode.}
When the predicted idle interval exceeds~$T_{\mathrm{enter}}$ and the
reachable SoC reduction exceeds a minimum useful
shift~$\Delta S_{\mathrm{min}}$, the target drops to
\begin{equation}
  S^*_{\mathrm{storage}}
  = \max\!\bigl(
      S_{\mathrm{idle}},\;
      S_{\mathrm{mid}} - \Delta S_{\max},\;
      S_{\mathrm{safe,min}}
    \bigr),
  \label{eq:storage_target}
\end{equation}
where
$\Delta S_{\max}
  = i^{\max}\,\max(0,\,T_{\mathrm{remain}} - T_{\mathrm{ready}}(S_{\mathrm{idle}}))
    \;/\;(\eta_d\,Q_{\max})$
and $T_{\mathrm{ready}}(S) = (S_{\mathrm{mid}} - S)\,Q_{\max}/(\eta_c\,i^{\max})$
is the time required to charge from~$S$ back to~$S_{\mathrm{mid}}$.
Because~$T_{\mathrm{remain}}$ decreases as the idle window elapses,
$S^*_{\mathrm{storage}}$ rises toward~$S_{\mathrm{mid}}$ automatically.
When $T_{\mathrm{remain}} < T_{\mathrm{ready}}(S_{\mathrm{current}})$,
the target reverts to~$S_{\mathrm{mid}}$.  In our prototype,
$T_{\mathrm{enter}} = 4$\,h and $\Delta S_{\mathrm{min}} = 0.02$.
 
\subsection{Inner Loop: Receding-Horizon QP}
\label{app:inner-loop}
 
Let $\mathcal{I} = (i_0,\dots,i_{H-1})$ be the corrective currents over
$H$~intervals of length~$\Delta t$, and let
$\mathcal{S} = (S_0,\dots,S_H)$ be the predicted SoC trajectory
initialized at the measured value~$\hat S_t$.  Define normalized
variables
\begin{equation}
  u_k = \frac{i_k}{i^{\max}},
  \qquad
  e_k = \frac{S_k - S^*}{\Delta S_{\mathrm{ref}}},
  \label{eq:norm_vars}
\end{equation}
where $\Delta S_{\mathrm{ref}} = S_{\mathrm{mid}} - S_{\mathrm{idle}}$.
 
The inner loop solves
\begin{align}
  \min_{\mathcal{I}}\;\;
  & \sum_{k=0}^{H-1}\!\Bigl[
      e_{k+1}^2
      + \lambda_I\, u_k^2
      + \lambda_\Delta\,(u_k - u_{k-1})^2
    \Bigr]
    + \lambda_T\, e_H^2
  \label{eq:qp_obj} \\[4pt]
  \text{s.t.}\;\;
  & S_{k+1} = S_k
    + \frac{\Delta t}{Q_{\max}}
      \bigl(\eta_c\,[i_k]^+ - \eta_d^{-1}\,[-i_k]^+\bigr),
  \label{eq:soc_dyn} \\
  & S_0 = \hat S_t,
  \label{eq:soc_init} \\
  & S_{\mathrm{safe,min}} \le S_k \le S_{\mathrm{safe,max}},
  \label{eq:soc_bounds} \\
  & |i_k| \le i^{\max},
  \label{eq:curr_bounds}
\end{align}
where $u_{-1}$ is the previously applied normalized current and
$[x]^+ = \max(x,0)$.  The controller applies only~$i_0$ and re-solves
at the next interval.  If $|\hat S_t - S^*| \le \varepsilon$, it sets
the current to zero.
 
The three ratios $\lambda_I$, $\lambda_\Delta$, $\lambda_T$ trade off
tracking speed against current magnitude and command smoothness.
We set them from two design targets: the desired correction timescale for
a representative SoC deviation, and the desired smoothness of the
maintenance-current trajectory.  The problem is a small convex QP,
feasible whenever $\hat S_t$ lies within hardware safe bounds.
\section{Software Components}

\subsection{GPU Burn Algorithm for Baseline}\label{app:software-burn}
\textbf{Calibration.} We first calibrate a tiny matrix--multiply kernel to learn a linear mapping between its duty cycle and GPU power, and then invert this mapping so we can interpolate for a target power and get back a duty cycle that achieves it. Here the \emph{duty cycle} $d \in [0,1]$ is the fraction of each fixed control window $T_{\text{win}}$ that the GPU spends actively running the GEMM kernel (for time $d \cdot T_{\text{win}}$) versus sleeping (for time $(1-d)\cdot T_{\text{win}}$), which smoothly scales the average power between idle ($d \approx 0$) and near-TDP ($d \approx 1$). Concretely, we run two small calibration tools on a single Titan~X: one sweeps over matrix sizes $N$ and duty cycles $d \in [0,1]$ using a duty-cycled GEMM loop in fixed windows and logs the resulting average GPU power to CSV, and the other uses the same GEMM burner to sweep only over $d$ for a fixed $N$. We then fit a simple linear model $P(d) \approx a d + b$ on the stable regime of the sweep and invert it to obtain $d(P)$ for our runtime ramps and checkpoint compensation.

\begin{algorithm}[t]
    \caption{Calibration of duty $\rightarrow$ power mapping}
    \label{alg:calibration}
    \begin{algorithmic}[1]
    \State Measure idle power $P_{\text{idle}}$ with GPU at rest
    \For{$N \in \mathcal{N}$} \Comment{matrix sizes}
      \State Allocate $A,B \in \mathbb{R}^{N \times N}$ on GPU
      \State Calibrate matmul time $\tau(N)$ using CUDA events
      \For{$d \in \mathcal{D}$} \Comment{duty cycles}
        \For{windows over fixed horizon}
          \State Run GEMMs for time $d \cdot T_{\text{win}}$ using $\tau(N)$
          \State Sleep for remaining $(1-d)\cdot T_{\text{win}}$
          \State Sample GPU power $P$ via NVML
        \EndFor
        \State Record $(N, d, \overline{P}, \overline{P} - P_{\text{idle}})$ to CSV
      \EndFor
    \EndFor
    \State Select fixed $N^\star$ and fit linear $P(d) \approx a d + b$ from CSV
    \State Define inverse mapping $d(P) = \mathrm{clip}\bigl((P-b)/a, 0, 1\bigr)$
    \end{algorithmic}
\end{algorithm}

\textbf{Integration in Training Loop.} In Algorithm \ref{alg:power-shaped-training} we show how we implement our GEMM burns during training. During warmup, before the first training step, we repeatedly run the kernel on both GPUs, gradually increasing the target power from a low “warmup” level to the normal training power over a fixed time window (e.g., 30 s). This creates a smooth ramp from idle to full load instead of a step change. The training loop is otherwise standard, except at each checkpoint. Since our GPUs are connected by NVLink, there is no need to communicate over a network and therefore the only dips we see are from the checkpointing itself. When rank 0 saves a checkpoint and its power drops, the other GPUs temporarily run the burn kernel at a higher target power chosen so that the sum of GPU power stays close to the normal training level. All ranks synchronize at a barrier before resuming training. After the last step, we run a symmetric cooldown: both GPUs gradually reduce their target power from the training level down to a lower “cool” level using the same burn kernel, again over a fixed time window. This produces a smooth ramp down to near-idle instead of a sudden drop.

\begin{algorithm}[t]
    \caption{GPU Burn Augmented Training}
    \label{alg:power-shaped-training}
    \begin{algorithmic}[1]
    \State  Calibrate linear map $P(d)$ and inverse $d(P)$ using GEMM burns
    \For{$t = 0$ to $T_{\text{warm}}$ step $\Delta t$} \Comment{warmup ramp}
      \State $P^\star \gets \text{lerp}(P_{\text{warm}}, P_{\text{train}}, t/T_{\text{warm}})$
      \State \Call{Burn}{$d(P^\star), \Delta t$}
    \EndFor
    \For{$s = 1$ to $S$} \Comment{training steps}
      \State \Call{TrainStep}{$s$}
      \If{$s \bmod K = 0$} \Comment{checkpoint every $K$ steps}
        \If{$\text{rank} = 0$}
          \State \Call{SaveCheckpoint}{$s$}
        \Else
          \State \Call{Compensate}{$P_{\text{train}}, P_{\text{ckpt}}$} \Comment{burn on rank $>0$}
        \EndIf
        \State \Call{CUDA\_Barrier()}{} \Comment{synchronize all ranks}
      \EndIf
    \EndFor
    \For{$t = 0$ to $T_{\text{cool}}$ step $\Delta t$} \Comment{cooldown ramp}
      \State $P^\star \gets \text{lerp}(P_{\text{train}}, P_{\text{cool}}, t/T_{\text{cool}})$
      \State \Call{Burn}{$d(P^\star), \Delta t$}
    \EndFor
    \end{algorithmic}
    \end{algorithm}

\textbf{Gloo and NCCL Barriers.} To make sure we can run checkpoint burn compensation without sacrificing training performance, we use a dual process group approach. We initialize two separate PyTorch distributed process groups: a primary NCCL group for all training communication (gradient synchronization, model updates), and a secondary Gloo group exclusively for checkpoint barriers. NCCL barriers enqueue operations on the CUDA stream, which blocks subsequent GPU kernels and prevents our GEMM burn from executing concurrently, something that Gloo barriers which use CPU-based synchronization primitives do not do. By routing only checkpoint synchronization through the Gloo group while maintaining NCCL for all training operations, we can achieve full NCCL training performance while allowing other GPUs to run compensation burns concurrently during checkpointing.
    
\end{document}